\documentclass[prx,twocolumn,english,superscriptaddress,floatfix,longbibliography]{revtex4-2}
\usepackage{amsmath}
\usepackage{mathtools}
\usepackage{float}
\usepackage{amsfonts}
\usepackage{amssymb}
\usepackage{dcolumn} 
\usepackage{array}
\newcolumntype{P}[1]{>{\centering\arraybackslash}p{#1}}
\newcolumntype{M}[1]{>{\centering\arraybackslash}m{#1}}
\newcolumntype{C}[1]{>{\centering\arraybackslwash}p{#1}}
\usepackage{float}
\usepackage{psfrag}
\usepackage{tabularx}
\usepackage{stackengine}
\usepackage{amssymb}
\usepackage{mathtools}  
\usepackage{xfrac} 
\usepackage[T1]{fontenc}
\usepackage{graphicx}
\usepackage{diagbox}
\usepackage[caption=false]{subfig}
\usepackage{tikz}
\usetikzlibrary{positioning}
\usetikzlibrary{arrows}
\usetikzlibrary{trees}
\usepackage{amssymb}
\usetikzlibrary{decorations.pathmorphing}
\usetikzlibrary{decorations.markings}
\usetikzlibrary{automata,positioning}
\usepackage{braket}
\usepackage{hyperref}
\usepackage{rotating}
\usepackage{adjustbox}
\usepackage{ragged2e}
\usepackage{simplewick}
\usepackage{simpler-wick}

\usepackage{csquotes}
\usepackage{physics,amsmath}
\usepackage{xcolor}
\usepackage{fancyhdr}
\UseRawInputEncoding
\usepackage[normalem]{ulem}

\usepackage{scalerel}

\begin{document}

\author{Sonaldeep Halder}
\affiliation{ Department of Chemistry,  \\ Indian Institute of Technology Bombay, \\ Powai, Mumbai 400076, India}

\author{Chayan Patra}
\affiliation{ Department of Chemistry,  \\ Indian Institute of Technology Bombay, \\ Powai, Mumbai 400076, India}

\author{Dibyendu Mondal}
\affiliation{ Department of Chemistry,  \\ Indian Institute of Technology Bombay, \\ Powai, Mumbai 400076, India}

\author{Rahul Maitra}
\email{rmaitra@chem.iitb.ac.in}
\affiliation{ Department of Chemistry,  \\ Indian Institute of Technology Bombay, \\ Powai, Mumbai 400076, India}

\title{Machine Learning Aided Dimensionality Reduction towards a Resource Efficient Projective Quantum Eigensolver}

\begin{abstract}
The recently developed Projective Quantum Eigensolver (PQE) has been demonstrated as an elegant methodology to compute the ground state energy of molecular systems in Noisy Intermdiate Scale Quantum (NISQ) devices. The iterative optimization of the ansatz parameters involves repeated construction of residues on a quantum device. The quintessential
pattern of the iteration dynamics, when projected as a time discrete map, suggests a hierarchical structure in the timescale of convergence,
effectively partitioning the parameters into two distinct classes. In this work, we have exploited
the collective interplay of these two sets of parameters via machine learning techniques to bring out the synergistic inter-relationship among them that triggers a drastic reduction in the number of
quantum measurements necessary for the parameter updates while maintaining the
characteristic accuracy of PQE. Furthermore the
machine learning model may be tuned to capture the noisy data of NISQ devices and thus the
predicted energy is shown to be resilient under a given noise model.

\end{abstract}

\maketitle

\section{Introduction}
The intractability in storing and manipulating many-body wavefunctions, which grow 
exponentially with the system size, restraints the 
application of classical many-body theories to large and highly correlated molecular systems. Quantum Computers offer an alternate infrastructure for simulating such systems\cite{PhysRevA.64.022319}, leveraging the principles of superposition and entanglement to orchestrate complex wavefunctions. The algorithms employing Quantum Phase Estimation (QPE) \cite{PhysRevLett.83.5162} to calculate the ground state energies of small molecular systems\cite{aspuru2005simulated} rely on the availability of 
fault-tolerant quantum devices with sufficient coherence time to indulge in their deep 
circuit requirements. Alternative hybrid algorithms like the popular Variational 
Quantum Eigensolver(VQE)\cite{Peruzzo2014}, Quantum Approximate Optimization 
Algorithm (QAOA)\cite{Farhi2014AQA}, Quantum Subspace 
Diagonalisation\cite{Motta2019, McClean2017,Stair2020,Huggins2020,https://doi.org/10.48550/arxiv.1909.08925}and Quantum Imaginary Time 
Evolution\cite{Motta2019,PRXQuantum.2.010317}  are more suitable for implementation on 
currently available Noisy Intermediate-Scale 
Quantum (NISQ)\cite{Preskill2018quantumcomputingin} devices which suffer from a variety of limitations,  
low coherence time, poor gate fidelity and readout errors being some of the prominent ones. The significance of VQE
in the NISQ era is primarily attributed to its exceptional noise resilience and the unprecedented level of versatility that it provides in terms of the choice of  
parameterized ansatz \cite{Szalay1995, Taube2006, Cooper2010, Evangelista2011, Harsha2018, Filip2020, Chen2021, Ryabinkin2018, Kandala2017} (which produces the trial wave function) and the availability of a wide range of classical numerical optimization techniques that can be employed to optimize the ansatz parameters. Despite its inherent robustness, VQE has limitations and shortcomings that impede its general applicability.
Foremost among these drawbacks are the issues of slow convergence and the tendency to become entrapped in barren plateaus where the gradient vanishes, rendering further progress impossible. Moreover, in the case of gradient-based optimization algorithms, the convergence process necessitates the calculation of a considerable number of energy gradients.

Projective Quantum Eigensolver (PQE)\cite{PRXQuantum.2.030301} 
offers a promising alternative approach which addresses some of the obstacles inherent to VQE without losing 
on the noise resilience characteristics. The methodology employed in this procedure involves the preparation of a trial state through the utilization of the disentangled unitary coupled cluster (dUCC) ansatz\cite{evangelista2019exact} which consists of an ordered 
product of parameterized unitary operators, each of which 
is obtained by exponentiating an excitation 
operator ($\hat{\kappa}_{\mu}$) (anti-hermitian analogue of 
classical cluster excitation operator in coupled cluster (CC) theory\cite{ek1966,vcivzek1969use,vcivzek1971correlation,bartlett2007coupled,crawford2000introduction}). The 
pool of excitation operators used for the ansatz  can 
either be selected adaptively or be built to include all 
unique excitations truncated up to a certain order. The 
optimization of the ansatz parameters is accomplished 
by minimizing the norm of the residue vector using a quasi-Newton type iterative 
procedure. This corresponds to the solution of the coupled multivariate
nonlinear equations fabricated by projecting the Schrodinger
equation by a selected set of many-body basis (which have 
one to one correspondence with the excitation operators 
included in the pool).

The feasibility (and cost) associated with the implementation 
of PQE on a quantum device is largely influenced by two major factors:
\begin{enumerate}
    \item The circuit depth corresponding to the realization of the required ansatz in terms of one and two qubit gates.
    \item The total number of measurements the PQE iterative procedure demands during the optimization of the ansatz parameters.
\end{enumerate}

One can utilize an appropriately prepared ansatz to address the issue concomitant with the depth of the implemented circuit. But even so, the cumulative quantum measurements required by the quasi-Newton iterative algorithm to determine the ground state energies translates to a protracted runtime on existing NISQ devices. Here we adopt an inter-disciplinary
approach to address this issue by borrowing the
pertinent ideas from nonlinear sciences and integrating them with the iterative optimization strategy with an aim to minimize the aggregate number of measurements that must be conducted during the PQE procedure.

It was previously shown by some of the present authors that
in the context of CC such nonlinear iterative optimization processes might be perceived
as an implicitly time-discrete multivariate map. The related set of 
coupled nonlinear equations that determine the parameters
undergoes period doubling bifurcation when their
optimization trajectory is perturbed beyond a critical 
point\cite{agarawal2020stability}. Such non-equilibrium phenomena follow
the universality of Feigenbaum dynamics\cite{feigenbaum1979universal,feigenbaum1978quantitative}
and as such, rationalizes the existence
of a set of collective modes consisting of a few of the \textit{dominant} (or \textit{principal})
amplitudes that macroscopically determine the optimization trajectory.
On the other hand, the \textit{recessive} (or \textit{auxiliary})
variables are enslaved, and as such, there exists a synchronization among these two sets of variables
during the optimization.
In general, the \textit{principal} and the \textit{auxiliary} modes 
have characteristically
different timescale of equilibration and
this temporal hierarchy may be used to
decouple these two sets of modes analytically. Moreover, 
the number of such \textit{principal}
modes are much less compared to the \textit{auxiliary} modes, whereas the former have significantly
larger magnitude than the latter. These typical characteristics license
us to employ the \textit{adiabatic approximation} and
\textit{Slaving Principle}\cite{synergetics1976introduction}
to freeze the time variation of the \textit{auxiliary} modes in the
characteristic timescale of the \textit{principal} modes
and ultimately express the \textit{auxiliary} variables as a function of
the \textit{principal} ones. The updated \textit{auxiliary} modes, however, are fed back to the
equations of the \textit{principal} modes, giving rise to a 
circularly causal relationship. By determining a functional dependence through which the \textit{principal}
amplitudes are mapped onto the \textit{auxiliary} amplitudes we can condense the problem only in
terms of the former\cite{agarawal2021accelerating,agarawal2021approximate,agarawal2022hybrid,patra2022synergistic}.

 In the current manuscript
this philosophy is rippled into the PQE framework where
such circular causality is exploited to effectively constrain the number of parameters that are freely optimizable in the description of the dUCC ansatz. A novel implementation is devised wherein only the \textit{principal} parameters are iteratively updated by means of measuring their corresponding residues, while the \textit{auxiliary} parameters are derived through the functional relationship that exists between these two sets of parameters.
To unravel this intricate relationship, we have astutely employed machine learning (ML) techniques, complete with an on-the-fly training mechanism. This ultimately leads to a dramatic reduction in the number of quantum measurements required for the iterative convergence of the ansatz parameters. We will refer our scheme as ML aided PQE(ML-PQE). In this context we must mention that certain other downfolding techniques have been recently developed
to solve classical coupled cluster theory using some sub-system embedding subalgebra (SES) techniques\cite{kowalski2018properties,bauman2019downfolding,bauman2019quantum,kowalski2020sub,bauman2022coupled} and later extended to
quantum computing framework as well\cite{metcalf2020resource,kowalski2021dimensionality,chladek2021variational,bauman2023coupled,mehendale2023parameter}.

In this article, we first provide the theoretical framework on PQE and the \textit{slaving principle}, leading to a new methodology for minimizing the number of measurements performed on a quantum computer. Then we move on to evaluate the efficacy and cost efficiency of our method on molecular systems.
This study is further extended to incorporate a stochastic error in the residues generated on a quantum device which eventually establishes the noise resilience of the developed method. In the last section we summarize the findings and discuss future research opportunities.


\section{Theory}
We divide the theory into three subsections : In subsection A, we briefly describe the basic theory and working equations used in PQE.
Subsection B mathematically investigates the relationship between the ansatz parameters and how it leads to a novel protocol for a cost effective PQE via the usage of  ML. In subsection C, details about the selected ML model is described briefly.

\subsection{Projective Quantum Eigensolver}
PQE involves the generation of a trial wavefunction $\ket{\Psi (\boldsymbol{\theta})}$ upon the action of a parameterized unitary ansatz
$\hat{U}(\boldsymbol{\theta})$ on a reference state $\ket{\Phi_o}$
\begin{equation} \label{psi}
   \ket{\Psi (\boldsymbol{\theta})} = \hat{U}(\boldsymbol{\theta})\ket{\Phi_o}.
\end{equation}

Incorporating $\ket{\Psi (\boldsymbol{\theta})}$ into the Schr\"{o}dinger
equation and employing the operation of left multiplication by $\hat{U}^{\dagger}(\boldsymbol{\theta})$, and subsequently projecting by
the excited determinants $\{\ket{\phi_{\mu}}\}$ leads to a set of residual equations 
\begin{equation} \label{4}
       r_{\mu}(\boldsymbol{\theta}) = \bra{\Phi_{\mu}}\hat{U}^{\dagger}(\boldsymbol{\theta})H \hat{U}(\theta)\ket{\Phi_o}
\end{equation}
The residual vector $r_\mu$ is minimized through
a classical optimizer such that $r_{\mu} \rightarrow 0$ (residual condition) 
and the corresponding PQE energy is computed as
\begin{equation} \label{3}
   \bra{\Phi_o}\hat{U}^{\dagger}(\boldsymbol{\theta})H \hat{U}(\boldsymbol{\theta})\ket{\Phi_o} = E_{PQE}(\boldsymbol{\theta}) \\
\end{equation}

Here, $\mu$ runs over the entire many-body basis (except the reference i.e $\mu \ne 0$). For cases where the number of ansatz parameters is lower than the number of many-body basis ($\{\Phi_{\mu}\}$), the residual condition can only be enforced for a subset of residues ($r_{\mu}$) which results in an uncertainty in the energy obtained by Eq. \eqref{3}\cite{PRXQuantum.2.030301}.

The PQE formalism has been built using the dUCC ansatz, which is characterized by a pool of ordered, non-commuting, anti-hermitian particle-hole operators ($\{\hat{\kappa}\}$) with the reference state taken to be the single Hartree-Fock (HF) determinant $\ket{\Phi_o} = \ket{\chi_i\chi_j...}$ , $\chi$s being the spin-orbitals. The ansatz can be written as

\begin{equation}\label{6}
    \hat{U}(\boldsymbol{\theta}) = \prod_{\mu} e^{\theta_{\mu}\hat{\kappa}_{\mu}}
\end{equation}

\begin{equation}
    \hat{\kappa}_{\mu}=\hat{\tau}_{\mu}-\hat{\tau}_{\mu}^\dagger
\end{equation}
    
\begin{equation}
    \hat{\tau}_{\mu}= \hat{a}^{\dagger}_{a}\hat{a}^{\dagger}_{b}....\hat{a}_{j}\hat{a}_{i}
\end{equation}

In the above equations, $\mu$ represents a multi-index particle-hole excitation structure as defined by the string of creation ($\hat{a}^{\dagger}$) and annihilation ($\hat{a}$) operators with the indices $\{i,j,...\}$ denoting the occupied spin-orbitals in the Hartree Fock state and $\{a,b,...\}$ denoting the unoccupied spin-orbitals. $\hat{\kappa}_{\mu}$ acts on the reference state to generate a many-body basis,

\begin{equation}
    \hat{\kappa}_{\mu}\ket{\Phi_o} = \ket{\Phi_{\mu}}
\end{equation}
For all practical applications, dUCC ansatz is constructed using only a subset of the total possible excitation operators ($\hat{\kappa}_{\mu}$) and the residual condition is implied over the corresponding basis.

The solution of the non-linear equations delineated by $r_{\mu}=0$ can be obtained numerically by employing the quasi-Newton iterative scheme:
\begin{equation} \label{iter}
    \theta_{\mu}^{(k+1)} = \theta_{\mu}^{(k)} + \frac{r_{\mu}^{(k)}}{D_{\mu}}
\end{equation}
where, $k$ represents an iteration level, $r_{\mu}$ denotes the residue, as defined by Eq. \eqref{4}, which is computed on a quantum device and the
denominator $D_{\mu}$ depends on the excitation structure of $\mu$ and is given as
($\epsilon_{i} + \epsilon_{j} + ...... - \epsilon_{a} - \epsilon_{b} - ....$) where
$\epsilon_p$ is the Hartree-Fock energy of the $p^{th}$ spin-orbital and $\{i,j,..\}$ are the orbital indices in Hartree-Fock determinant
$\ket{\Phi_o}$ which has been replaced by orbitals of indices $\{a,b,...\}$ to form the excitation structure of $\mu$. Usually, the iterative
scheme is deemed converged when the \textit{2-norm} of the residue vector ($\| r\|$) goes below a predefined threshold (usually taken to be a small value
like $10^{-05}$). The expression,  $r_{\mu} = \bra{\Phi_{\mu}}\Bar{H}\ket{\Phi_o}$, where
$\Bar{H}=\hat{U}^{\dagger}(\theta)\hat{H}\hat{U}(\theta))$, can be  further
broken down into diagonal terms.

\begin{equation} \label{12}
    r_{\mu} = \bra{\Omega_{\mu}(\frac{\pi}{4})}\Bar{H}\ket{\Omega_{\mu}(\frac{\pi}{4})} -\frac{E_{\mu}}{2} - \frac{E_o}{2}
\end{equation}
Here, $E_{\mu} = \bra{\Phi_{\mu}}\Bar{H}\ket{\Phi_{\mu}}$, $E_o = \bra{\Phi_o}\Bar{H}\ket{\Phi_o}$ and $\ket{\Omega_{\mu}(\frac{\pi}{4})}=e^{\frac{\pi}{4}\hat{\kappa}_{\mu}}\ket{\Phi_o}$. Equation \eqref{12} enables us to calculate the residues only by measuring the diagonal terms on a quantum computer. Thus, Eq. \eqref{iter} in conjunction with Eq. \eqref{12} gives a complete recipe for the iterative optimization of the ansatz parameters in the quantum computing framework.

\subsection{A Brief Account of the General Mathematical Structure of the Slaving Principle and the Algorithmic Development Towards ML-PQE:}

In many practical scenarios one is confronted with multi-variable dynamical systems having a collective motion of
co-existing fast and slow relaxing modes, manifesting a temporal hierarchy among the constituent variables.
The long time behaviour of the dynamics is solely dictated by the slow relaxing modes \cite{Haken1983} which are often identified by the
\textit{linear stability analysis}\cite{strogatz2018nonlinear,alligood1996two,eubank1997introduction}.
While the \textit{linear stability analysis} provides a 
robust choice, the construction and the diagonalization of 
the \textit{stability matrix} is often a computationally expensive step. For many of
the practical cases this can be bypassed by judiciously 
choosing the \enquote{important} set of parameters which mimic the
phase space trajectory of that shown by the full set of 
parameters. Consequently, if we confine our attention only to these \enquote{important} parameters,
the effective dimensionality
of the system is significantly reduced and can be exploited to gain enormous computational advantages.

\begin{figure*}[!ht]
    \centering
\includegraphics[width=0.9\linewidth]{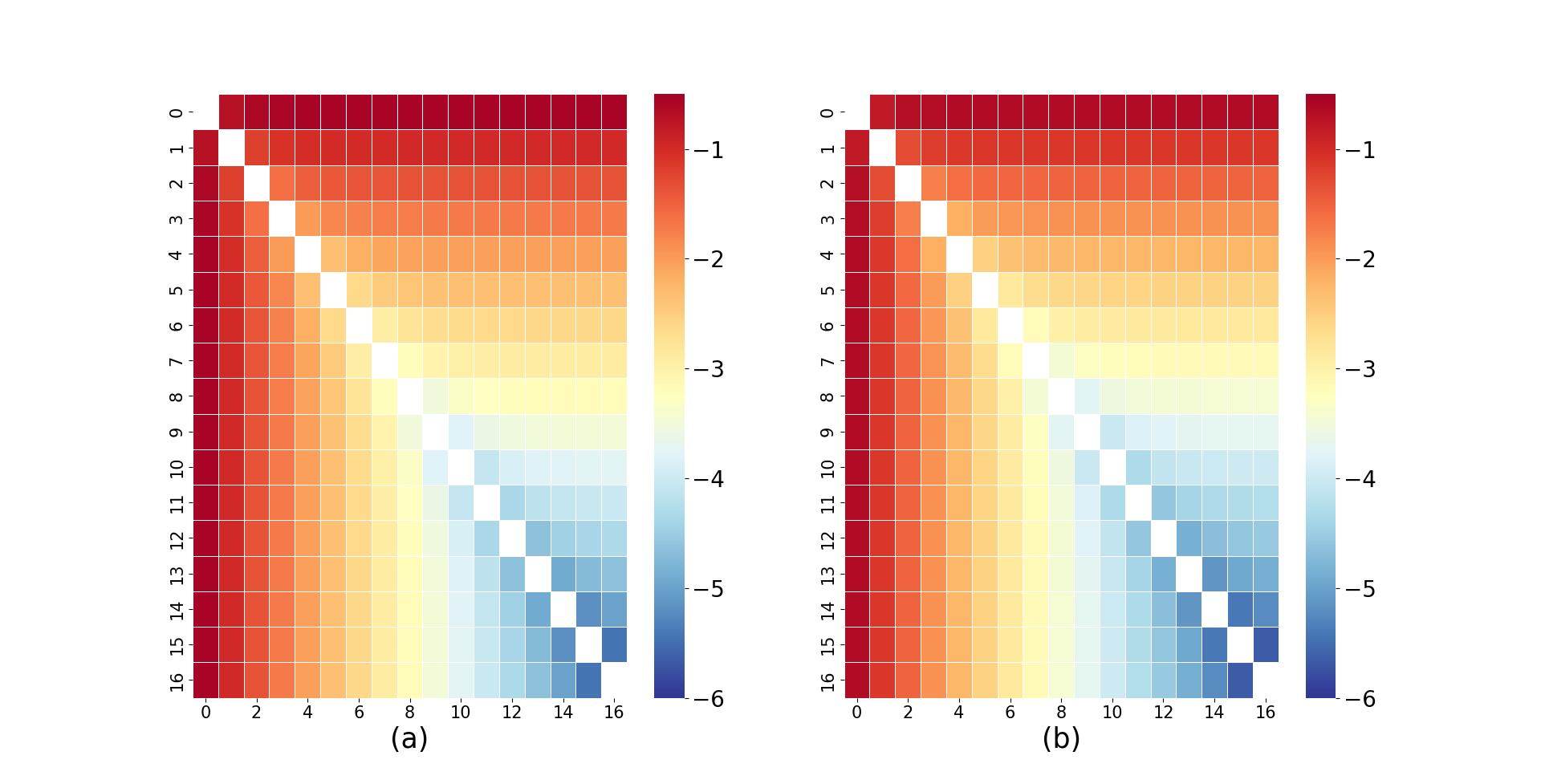}
\caption{A comparison of the Distance Matrix (a) Constructed using all the parameters and (b) Constructed using $20\%$ relatively large (absolute values after the last iteration) parameters for $H_2O$ $(\angle H\mbox{-}O\mbox{-}H = 104.4776^o, r_{O-H}=0.958 \mbox{\normalfont\AA})$ in logarithmic scale. }
    \label{fig: heatmap}
\end{figure*}

A study of the PQE convergence trajectory encompassed during the quasi-Newton iterative procedure reveals a similar underlying architecture. An illustrative insight to this can be obtained by plotting the distance matrix with element $M_{ij}^S$
corresponding to the \textit{2-norm}  $\|\theta_x^S-\theta_y^S\|$. Here, $S$ indicates the dimension of the parameter space taken during the construction of the matrix and $\theta_x^S, \theta_y^S$ are the parameter vectors at $x^{th}$ and $y^{th}$ iteration. 
The distance matrix gives a comprehensive description of the phase space trajectory of the entire iterative process highlighting
its dynamical behaviour.
As evident from Fig. \ref{fig: heatmap}, the distance matrix constructed by taking a small fraction (about $20\%$) of parameters that have relatively larger absolute magnitudes is nearly
identical to that constructed using the entire parameter set.
In other words, it is only these small number of parameters chosen in the described manner that contribute in a major capacity towards the phase space trajectory traversed during the PQE iterative scheme.
This particular behaviour corroborates the existence of two distinct classes of
parameters present\cite{agarawal2021accelerating,agarawal2021approximate,patra2022synergistic}
\begin{enumerate}
    \item Principal Parameter Subset (\textit{PPS}): A relatively smaller subset $\{\theta_P\}$
    consisting of parameters that have larger absolute magnitudes and longer relaxation timescale to reach the fixed point.
    They are the ones which dictate the optimal set.
    \item Auxiliary Parameter Subset (\textit{APS}): A relatively larger subset $\{\theta_A\}$ consisting of parameters that have
    smaller absolute magnitudes and much smaller relaxation timescale of convergence.
\end{enumerate}
Based on this observation, our motive is to decouple the parameter
space into \textit{PPS} and \textit{APS} and reduce the dimensionality using the \textit{slaving principle}
to devise a cost-effective variant of PQE.

 Towards the iterative determination
of the parameters, one may start with Eq. \ref{iter} which are analyzed as a multivariate time-discrete map
where each iteration is taken to be a unit time-step.
The residue vector $r_\mu$ may be expanded
via the Baker-Campbell-Hausdorff multi-commutator expansion:
\begin{equation} \label{theta_mu}
\begin{split}
 \Delta \theta_{\mu} D_{\mu} = r_{\mu}& = \bra{\Phi_{\mu}} (\prod_{\lambda} e^{\theta_{\lambda}\hat{\kappa}_{\lambda}})^\dagger \hat{H} \prod_{\nu} e^{\theta_{\nu}\hat{\kappa}_{\nu}} \ket{\Phi_0} \\
 & = \bra{\Phi_{\mu}} \hat{H} + \sum_{\nu} \theta_{\nu} [\hat{H},\hat{\kappa}_{\nu}] \\
 & + \sum_{\nu} \sum_{\lambda > \nu} \theta_{\nu}
 \theta_{\lambda} \Big[[H,\hat{\kappa}_{\nu}],\hat{\kappa}_{\lambda}\Big] + . . . \ket{\Phi_0}
\end{split}
\end{equation}
where, $\Delta{\theta_{\mu}} = \theta_{\mu}^{(k+1)} - \theta_{\mu}^{(k)}$. The operator \enquote{$\Delta$} can be viewed as the
discrete time analogue of the time derivative operator. As already discussed, the distribution of magnitudes and relaxation timescale allow us
to decompose these parameters ($\theta_\mu$) into $\{\theta_P\}$
with number of elements $n_P$ and $\{\theta_A\}$ with number
of elements $n_A$ with the condition $n_A >> n_P$.
The collective index of excitation associated with each element of \textit{PPS} is
to be denoted as $I$ whereas
those associated with the \textit{APS} will be denoted by $i$. Note that for both set of
amplitudes, Eq. \eqref{theta_mu} holds. In order to simplify the mathematical 
manipulations to express the \textit{auxiliary} amplitudes in terms of the \textit{principal} 
amplitudes, one may further extract out the diagonal part from the linear 
term for \textit{APS}:
\begin{equation} \label{theta_L extracted diagonal}
    \begin{split}
        & \Delta \theta_{A_i} = \frac{1}{D_{A_i}} \bra{\Phi_{A_i}} \hat{H} + \theta_{A_i} [\hat{H},\hat{\kappa}_{A_i}] + \sum_{\nu \neq A_i} \theta_{\nu} [\hat{H},\hat{\kappa}_{\nu}] \\
 & + \sum_{\nu} \sum_{\lambda > \nu} \theta_{\nu}
 \theta_{\lambda} \Big[[H,\hat{\kappa}_{\nu}],\hat{\kappa}_{\lambda}\Big] + . . . \ket{\Phi_0}
    \end{split}
\end{equation}
where the \textit{diagonal term} is taken out of the first commutator. 
Eq. \eqref{theta_L extracted diagonal} can be put to a compact form
\begin{equation} \label{del_thetaS}
\begin{split}
    \Delta \theta_{A_i} &=  \lambda_{A_i} \theta_{A_i} + M_{A_i}(\{\theta_P,\theta_A\}); \forall i= 1, 2, ..., n_A
\end{split}
\end{equation}
with the defintion
\begin{equation}
\begin{split}
    & \lambda_{A_i} = \frac{1}{D_{A_i}} \bra{\Phi_{A_i}}  [\hat{H},\hat{\kappa}_{A_i}] \ket{\Phi_{0}}
\end{split}
\end{equation}

In a similar fashion the corresponding equations for the \textit{PPS} can also be written
\begin{equation} \label{del_tL}
\begin{split}
    \Delta \theta_{P_I} &= \lambda_{P_I} \theta_{P_I}+ Q_{P_I}(\{\theta_{P},\theta_{A}\}); \forall I= 1, 2, ..., n_P
\end{split}
\end{equation}
Here, $Q_{P_I}(\{\theta_{P},\theta_{A}\})$ and $M_{A_i}(\{\theta_P,\theta_A\})$ contain all the nonlinear,
off-diagonal and coupling terms:
\begin{equation} \label{P}
    \begin{split}
        &M_{A_i}(\{\theta_{P},\theta_{A}\}) = \frac{1}{D_{A_i}} \bra{\Phi_{A_i}} \hat{H} + \sum_{\nu \neq A_i} \theta_{\nu} [\hat{H},\hat{\kappa}_{\nu}] \\
 & + \sum_{\nu} \sum_{\lambda > \nu} \theta_{\nu}
 \theta_{\lambda} \Big[[H,\hat{\kappa}_{\nu}],\hat{\kappa}_{\lambda}\Big] + . . . \ket{\Phi_0}
    \end{split}
\end{equation}

\begin{equation} \label{Q}
    \begin{split}
        &Q_{P_I}(\{\theta_{P},\theta_{A}\}) = \frac{1}{D_{P_I}} \bra{\Phi_{P_I}} \hat{H} + \sum_{\nu \neq P_I} \theta_{\nu} [\hat{H},\hat{\kappa}_{\nu}] \\
 & + \sum_{\nu} \sum_{\lambda > \nu} \theta_{\nu}
 \theta_{\lambda} \Big[[H,\hat{\kappa}_{\nu}],\hat{\kappa}_{\lambda}\Big] + . . . \ket{\Phi_0}
    \end{split}
\end{equation}

Note that in both the equations above, we have not yet imposed any restrictions on the labels $\mu,\nu$.
To obtain the most general solution for the \textit{auxiliary} parameters one can reorganize Eq. \eqref{del_thetaS} as
\begin{equation} \label{reorg_del_thetaS}
\begin{split}
    (\Delta - \lambda_{A_i}) & \theta_{A_i} = M_{A_i}(\{\theta_P,\theta_A\})\\
    \implies & \theta_{A_i} = (\Delta - \lambda_{A_i})^{-1} M_{A_i}(\{\theta_P,\theta_A\}) 
\end{split}
\end{equation}

The inverse operator $(\Delta - \lambda_{A_i})^{-1}$ leads to a power series expansion of Eq. \eqref{reorg_del_thetaS}\cite{haken1997discrete,haken1982slaving}
\begin{equation}\label{summed_thetaSi}
    \theta_{A_i} = \sum_{m=-\infty}^{l} (1+ \lambda_{A_i})^{l-m} M_{A_i}(\{\theta_P,\theta_A\})
\end{equation}
Applying \textit{summation by parts}, a mathematical technique similar to \textit{integration by parts},
Eq. \eqref{summed_thetaSi} can be further expanded

\begin{equation} \label{general_thetaS} 
\begin{split}
  & \theta_{A_i} = M_{A_i} \sum_{m=-\infty}^{l} (1+ \lambda_{A_i})^{l-m} \\ 
  & - \sum_{m=-\infty}^{l} (1+ \lambda_{A_i})^{l+1-m} \sum_{m'=-\infty}^{m-1} (1+ \lambda_{A_i})^{m-1-m'} \Delta M_{A_i} 
\end{split}
\end{equation}

The summations in equation Eq. \eqref{general_thetaS} can be readily performed, resulting in the final expression\cite{patra2022synergistic,haken1997discrete,haken1982slaving,wunderlin1981generalized}

\begin{equation} \label{thetaS_ad+p_ad}
    \theta_{A_i} = {-\frac{M_{A_i}(\{\theta_P,\theta_A\})}{\lambda_{A_i}}} {-\frac{\Delta M_{A_i}(\{\theta_P,\theta_A\})}{\lambda_{A_i}^2}} 
\end{equation}

\begin{figure*}[!ht]
    \centering
\fbox{\includegraphics[width=\textwidth]{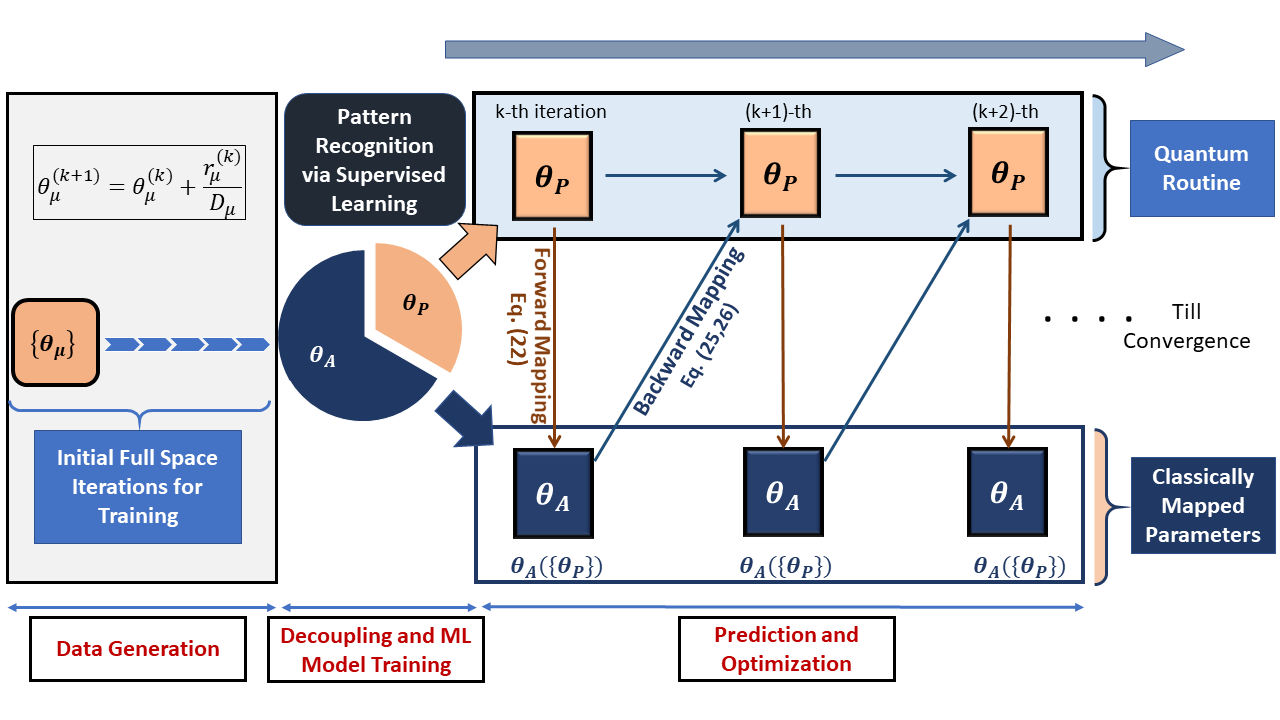}}
\caption{A schematic representation of the algorithm where the \textit{auxiliary} parameters, $\theta_A$ are determined through unique function of the \textit{principal} parameters, $\theta_P$, which are in turn updated through the back coupling of both sets of amplitudes.
The overall algorithm utilizes significantly less quantum resources as it decreases the dependence on the quantum sub-part for
residue construction.}
    \label{fig:iteration}
\end{figure*}
The first term in
Eq. \eqref{thetaS_ad+p_ad} may be called the adiabatically approximated term as it can be obtained alternatively by
simply putting $\Delta \theta_{A_i}=0$ in Eq. \eqref{del_thetaS} as
the \textit{adiabatic approximation}\cite{synergetics1976introduction} suggests. To obtain a fully decoupled set of equations only for the
\textit{auxiliary} variables we can approximately neglect\cite{haken1977chapter,agarawal2021approximate,patra2022synergistic} all the $\theta_A$ that are present in $M(\{\theta_P,\theta_A\})$
with respect to $\theta_P$ due to their relative magnitude -

\begin{equation} \label{thetaS as only thetaL}
    \begin{split}
        \theta_{A_i} &= -\frac{M_{A_i}(\{\theta_A,\theta_P\})}{\lambda_{A_i}}-\frac{\Delta{M(\{\theta_P,\theta_A\})}}{{\lambda_{A_i}}^2}\\
        & \xrightarrow{{\mid \theta_A \mid } \approx 0} -\frac{M_{A_i}(\{\theta_P\})}{\lambda_{A_i}} -\frac{{M(\{\theta_P,\Delta{\theta_P}\})}}{{\lambda_{A_i}}^2}
    \end{split}
\end{equation}

With the adiabatic and decoupling conditions, in Eq. \eqref{thetaS as only thetaL} the linearity of $\Delta$ operator ensures that the higher order
post-adiabatic terms depend on both \textit{principal} parameters and their successive differences (i.e. $\Delta{\theta_P}$).
While one can dive deep into further mathematical jargons of Eq. \eqref{thetaS as only thetaL}, the essential idea
we want to establish here is that in principle the \textit{auxiliary} amplitudes can be expressed solely in terms
of the \textit{principal} amplitudes and their successive time difference i.e. 
\begin{equation} \label{functional generic rel.}
    \theta_A = f(\theta_P,\Delta{\theta_P})
\end{equation}

To manifest the synergy between the two classes of variables, the \textit{auxiliary} parameters determined at $k$-th
step are fed back to the dynamics of the \textit{principal} ones via the coupling present in $Q$ (see Eq. \eqref{Q})
in the next $(k+1)$-th iteration and this \textit{circular causality loop} continues till the fixed point is reached.
Note here, despite the fact that the \textit{adiabatic approximation} results in the
temporary immobilization of \textit{auxiliary} parameters
during each iteration step, they continue to evolve dynamically due to their parametric
dependence on the \textit{principal} modes, thereby constantly
changing from one step to the next.
In this manner, effectively the overall dynamics is governed by the \textit{principal} parameters only, while
the \textit{auxiliary} parameters linger around as a function of the former. This dynamical interplay
substantially reduces the effective \textit{degrees of freedom} of the parameter space. It can be quickly realized that application of this modus operandi to the PQE framework can tremendously cut down the number of independent parameters which are optimized using quantum resources and thus potentially curtails the number of measurements.

While finding out the exact analytical structure of the intricate
inter-relationship between the two classes of variables is a stimulating challenge, in this
work we resort
to ML to map the \textit{auxiliary} parameters
from the \textit{principal} parameters $(f : \{\theta_P, \Delta{\theta_P}\} \rightarrow \{\theta_A\})$ by recognizing the quintessential pattern
that exists among them.
Below we provide a road-map to such a prototypical ML algorithm
\begin{enumerate}
    \item Use the conventional PQE and iterate for $n$ steps while recording the data associated with the evolution of the parameters.
    \item Train and deploy a  ML model:
    \begin{itemize}
        \item  Label the parameters as \enquote{\textit{principal}} and \enquote{\textit{auxiliary}} based on their absolute magnitudes, obtained after the $n^{th}$ iteration.
        \item Segregate the saved data into feature and target variables based on the labels obtained in the previous step 
        \item Train an appropriate ML model on the segregated data and deploy it for generating predictions.
    \end{itemize}
    \item Execute the subsequent iteration by solely updating the parameters labeled as \textit{principal} via residue construction and predict the remaining ones using the trained ML model. Collate all the parameters and employ them in the ensuing iteration to once again update the \textit{principal} parameters. This constitutes what is known as the feedback mechanism. Keep iterating in this described manner till the convergence is reached.
\end{enumerate}

We can view the last step described above in terms of the ansatz. Once the ML model is trained, the dUCC ansatz

\begin{equation}\label{non-ml-ansatz}
    \begin{split}
        \hat{U}(\boldsymbol{\theta}) &= \prod_{\mu}e^{\theta_{\mu}\hat{\kappa}_{\mu}}\\
        &=e^{\theta_{1}\hat{\kappa}_{1}}e^{\theta_{2}\hat{\kappa}_{2}} \dots e^{\theta_{k}\hat{\kappa}_{k}} \dots e^{\theta_{l}\hat{\kappa}_{l}}   \dots \\ &e^{\theta_{q}\hat{\kappa}_{q}} \dots e^{\theta_{\scaleto{N_{par}}{3.8pt}}\hat{\kappa}_{\scaleto{N_{par}}{3.8pt}}}
    \end{split}
\end{equation}
where, $N_{par}$ represents total number of parameters, gets dynamically mapped to a form where only $n_P$ \textit{principal} parameters are treated as independent whereas $n_A$ \textit{auxiliary} parameters are expressed in terms of the \textit{principal} ones:

\begin{equation}\label{ml-ansatz}
    \begin{split}
        \hat{U}_{ML}(\boldsymbol{\theta_P}) &= e^{\theta_{\scaleto{P_1}{3.8pt}}\hat{\kappa}_{\scaleto{P_1}{3.8pt}}}e^{\theta_{\scaleto{P_2}{3.8pt}}\hat{\kappa}_{\scaleto{P_2}{3.8pt}}} \dots e^{\theta_{\scaleto{A_{k}}{3.8pt}}^{\scaleto{ML}{3.8pt}}\{\theta_P\}\hat{\kappa}_{\scaleto{A_k}{3.8pt}}} \dots\\ 
        & e^{\theta_{\scaleto{A_{l}}{3.8pt}}^{\scaleto{ML}{3.8pt}}\{\theta_P\}\hat{\kappa}_{\scaleto{A_l}{3.8pt}}} \dots e^{\theta_{\scaleto{A_{n_A}}{3.8pt}}^{\scaleto{ML}{3.8pt}}\{\theta_P\}\hat{\kappa}_{\scaleto{A_{n_A}}{3.8pt}}} \dots \\ &e^{\theta_{\scaleto{P_{n_P}}{3.8pt}}\hat{\kappa}_{\scaleto{P_{n_P}}{3.8pt}}}
    \end{split}
\end{equation}

The order of the exponential operators remains the same for the ML ($\hat{U}_{ML}$) and 
the standard ansatz ($\hat{U}$). However, which of the parameters are eventually labeled as 
\textit{auxiliary} or \textit{principal} depends on the particular problem and in this sense, the ML enabled 
ansatz is dynamic in nature. Eq. \eqref{ml-ansatz} represents a generic case where 
$\theta_k$, $\theta_l$ and $\theta_q$ get labeled as \textit{auxiliary} (and are now represented as 
$\theta_{A_k}^{ML}$, $\theta_{A_l}^{ML}$ and $\theta_{A_{n_A}}^{ML}$).
It is evident from this form of the ansatz that the parameter update equation (Eq. \eqref{iter}) now needs to be applied only for updating $\{\theta_P\}$:
\begin{equation}
    \theta_{P_I}^{(k+1)} = \theta_{P_I}^{(k+1)} + \frac{r^{(k)}_{P_I}}{D_{P_I}}
\end{equation}
where $r_\mu$ is now constructed as the 
effective hamiltonian matrix elements between the HF function
and a set of selected excited determinants which are generated by
the action of \textit{principal} excitation operators ($\hat{\kappa}_{P_I}$) on the HF function.
\begin{equation} \label{reduced-iter}
    r_{P_I} = \bra{\Phi_{P_I}}\hat{U}^{\dagger}_{ML}(\boldsymbol{\theta_P})H\hat{U}_{ML}(\boldsymbol{\theta_P})\ket{\Phi_o}
\end{equation}
The restriction on the projective subspace leads to a
significant reduction in the number of measurements that need 
to be carried out on the quantum platform.
 A bird's eye-view of the entire optimization process is presented in Fig. \ref{fig:iteration}.

For conventional PQE with $N_{par}$ ansatz parameters, the upper bound to the number of measurements ($m_{res}$)  required to construct the residue vector with the \textit{2-norm} being within the precision of $\Bar{\epsilon}_{res}$ is given by\cite{PRXQuantum.2.030301}

\begin{equation}\label{total_measurement}
    m_{res} \le 3(N_{par})\frac{(\Sigma_{l}|h_l|)^2}{\Bar{\epsilon}_{res}^2},
\end{equation}
where $h_l$ represents the coefficient of the $l^{th}$ Pauli-string obtained after transforming the Hamiltonian. If during the entire iterative convergence procedure, the number of times the residue vector is constructed is given as $N_{res}^{PQE}$, the corresponding total number of measurements ($m_{PQE}$) is bounded by

\begin{equation}
    m_{PQE} \le N_{res}^{PQE}m_{res} 
\end{equation}

The training phase of ML-PQE utilizes the same number of measurements as that of the conventional one. But once the training is complete, the ML embedded ansatz (Eq. \ref{ml-ansatz}) comes into play. The measurements are performed only to update the \textit{principal} parameters whereas the remaining parameters are predicted using the trained ML model. The number of measurements ($m_{res}^{ML}$) required to form the reduced residue vector is bounded as

\begin{equation}\label{reduced_measurement}
    m_{res}^{ML} \le 3 (n_P) \frac{(\Sigma_{l}|h_l|)^2}{\Bar{\epsilon}_{res}^2}
\end{equation}

As $n_p$ is only a tiny fraction of $N_{par}$ ($n_p \ll N_{par}$), the upper bound for the number of measurements required post training in the ML-PQE scheme is substantially lower. 
The total number of measurements ($m_{PQE}^{ML}$) required for the ML-PQE scheme, including the training iterations, can be depicted as

\begin{equation}
    m_{PQE}^{ML} \le (N_{res}^{PQE-Tr}m_{res} + N_{res}^{PQE-Pr}m_{res}^{ML})
\end{equation}
where, $N_{res}^{PQE-Tr}$ denotes the number of times the residue vector is constructed during training and $N_{res}^{PQE-Pr}$, the number of residue vectors constructed thereafter. It is to be noted that if the ML model works properly, 
\begin{equation}
N_{res}^{PQE-Tr} + N_{res}^{PQE-Pr} = N_{res}^{ML-PQE} \lesssim N_{res}^{PQE}
\end{equation}
which can be verified later in the results and discussions section. Now, as $m_{res}^{ML} \ll m_{res}$, it implies
$m_{PQE}^{ML} \ll m_{PQE}$.
Although a smaller dimension for the \textit{PPS} is preferable to reduce the number of measurements, the ability of the ML model to accurately map
the \textit{PPS} to \textit{APS} is also an important consideration. A lower dimensional space may still reproduce the trajectory but the
restricted performance of the ML model demands for an optimally chosen \textit{PPS} size. Apart from the cost reduction, one must also focus on the
accuracy obtained using the ML-PQE scheme. Since the ML model only approximates the functional dependence through training, there is an inherent
error that is expected. But as is apparent in the later sections, this error is too low to have any practical altercations.

\subsection{Details of the Machine Learning Model}
The utilization of ML methodologies has become increasingly ubiquitous across various domains of science and technology, aimed at
discovering insightful patterns within the given data. Within the context of this study, our objective is to capture the latent relationship
between \textit{PPS} and \textit{APS}, which  perfectly aligns with the abilities offered by ML. Given the inherent nature of our problem,
entailing the prediction of continuous target values, a regression-based ML model was deemed suitable for our purposes. To this end,
we employed the Kernel Ridge Regression (KRR) model, which approximates a function as

\begin{equation}
\begin{split}
\theta_A^i(w,\boldsymbol{\theta_P}) &= w_o^i + w_1^i\theta_{P_1} + w_2^i\theta_{P_2} + ..... + w_p^i\theta_{P_{n_P}} \\
& =W_i^T \Theta_P
\end{split}
\end{equation}
where $W_i = [w_o^i, w_1^i, w_2^i, ....]$ and $\Theta_P = [1, \theta_{P_1}, \theta_{P_2}, \theta_{P_3}, ....]$ are column matrices. The optimized weights, $\{w\}$, are determined via the minimization of cost function (J) with respect to the weight vector W 

\begin{equation}
    J_i(W) = \sum_{m} \|W^T \Theta_P^m - \theta_{A_m}^i\|^2 + \alpha\|W'\|^2
\end{equation}
Here, $W'=[w_1^i, w_2^i, ....]$, $\Theta_P^m$ is the segregated \textit{principal} parameter vector at $m^{th}$ training set and $\theta_{A_m}^i$ is the value of $i^{th}$ \textit{auxiliary} parameter in the  $m^{th}$ training set. The complexity parameter, $\alpha$, is used to fine-tune the regularization strength of the model to combat overfitting.

A linear model, as the one used here, can learn the inherent non-linear structure of the function connecting the \textit{principal} and \textit{auxiliary} parameters once the original features have been transformed to a higher dimensional space. However, the process of determining the optimal weights in such cases can prove to be prohibitively expensive. Fortunately, there exists a technique known as the kernel trick, which can effectively circumvent this issue by avoiding the explicit transformation of the feature space and instead relying on kernel functions to directly compute the necessary inner products between the transformed features. In this manuscript, we use the popular radial basis function (RBF) kernel\cite{scikit-learn}, which has been found to adequately capture the non-linear signature between the said 
two classes of parameters.

In order to effectively detect the intricate underlying relationships between variables, it is typically necessary to procure a large training data set.
However, such an undertaking necessitates the utilization of a significant quantum resources, which runs counter to the objective of minimizing resource usage.
Thus, a methodology must be devised that secures only the requisite amount of data that is sufficient for the accurate training of the model.
In the present investigation, this goal is achieved by imposing a threshold, which we have coined as the Limiting Residue-Norm Threshold (LRNT), on the norm of the residue ($\|r_{\mu}\|$) during the iterations carried out over the full space. 
Put simply, the training process is terminated when the \textit{2-norm} of the residue falls below the LRNT value (i.e., $\|r_{\mu}\| \le LRNT$).
Our empirical findings indicate that setting the LRNT value within the range of 0.02 to 0.005 yields a sufficiently large training data set for the selected model. Upon the completion of the training process, the overall form of the ansatz is described by Eq. \eqref{ml-ansatz}. The iterative procedure is implemented by leveraging Eq. \eqref{reduced-iter} in conjunction with the predictions generated by the ML model until convergence is achieved, which is marked by norm of the reduced residue vector going below a certain small threshold.

\section{Results and Discussion}

\begin{figure*}[!ht]
    \centering
\includegraphics[width=0.9\textwidth]{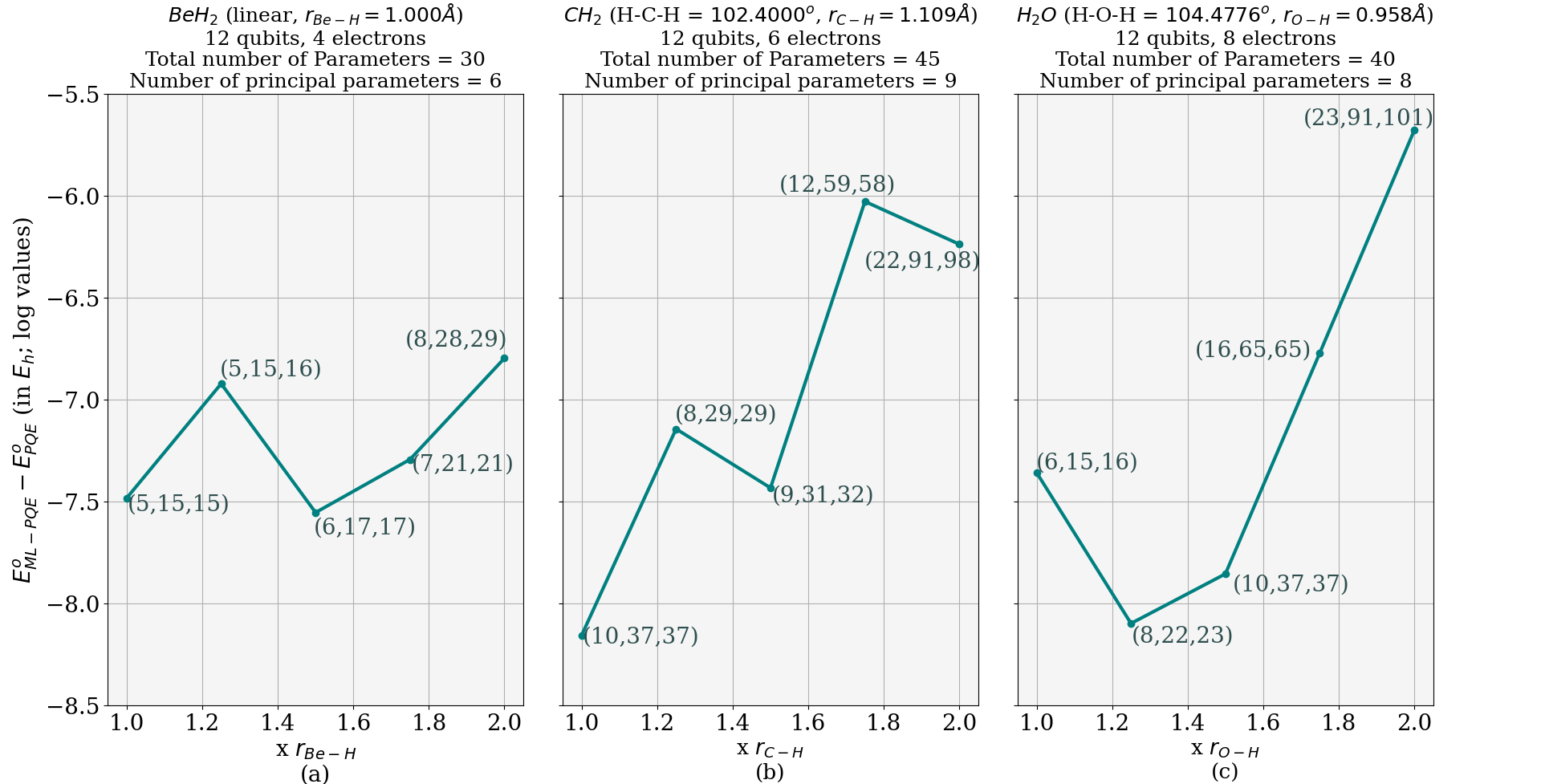}
\caption{Accuracy plot for (a)$BeH_2$, (b)$CH_2$ and (c)$H_2O$ showcasing the difference between the converged energies obtained using ML-PQE ($E^o_{ML-PQE}$) and that using conventional PQE ($E^o_{PQE}$) at different geometries. dUCCSD ansatz in used for all the cases with the total number of parameters mentioned atop each plot. Each point is annoted by (number of training set, total iterations for ML-PQE, total iterations for conventional PQE).}
    \label{fig:accuracy}
\end{figure*}

Through this section, our primary objective is to conduct a comparative analysis between the accuracy achieved by the ML-PQE protocol and the conventional one while also taking into consideration the associated cost. Before moving forward, it is imperative to acknowledge that the set of excitation operators ($\{\hat{\kappa_{\mu}}\}$) used in the 
construction of $\hat{U}$ and the corresponding $\hat{U}_{ML}$ are identical and as such, the 
latter provides no additional expressibility\cite{sim2019expressibility}. 
Hence, this study aims to assess the ability of ML-PQE to replicate the outcomes of its conventional counterpart, while using only a restricted amount of quantum computing resources. Additionally, we examine the performance of the newly developed protocol under the Gaussian noise model to determine its viability for NISQ platforms. To this end, we elucidate the utilization of the ML model's integrated regularization mechanism to counteract overfitting that is likely to manifest in data procured from an actual, noisy quantum device. We divide this study into two parts: 
in subsection A, we use the ML-PQE ansatz ($\hat{U}_{ML}$) to calculate the ground state 
energies of three molecular systems viz $BeH_2$, $CH_2$ and $H_2O$ over a range of their 
geometries and compare them with that obtained by conventional PQE. In this subsection, 
we also highlight the effective reduction in the iteration cost offered by the ML-PQE protocol. 
Subsection B is dedicated to analyzing the effect of Gaussian error on the convergence 
trajectory of ML-PQE and its comparison with the conventional one.

For all calculations, we have used the orbitals obtained from the restricted Hartree-Fock method
provided by PySCF\cite{sun2018pyscf} using the STO-3G basis set. The required Jordan-Wigner transformation for the
Hamiltonian and the ansatz is provided by the qiskit-nature\cite{Qiskit} modules. All simulations have been 
performed on the \textit{statevector simulator} provided by qiskit. Since our study does not 
concern the optimal ordering of the operator pool, we have used the default one provided by 
the qiskit's UCC class. Moreover, no supplementary convergence acceleration techniques, such as Direct Inversion in the Iterative Subspace (DIIS) were used. The ML pipeline consists of the usual procedure of standardizing the features (by subtracting the mean and scaling to unit variance) prior to the training and prediction stages.

\subsection{Accuracy of ML-PQE Ansatz and Its Associated Cost Consideration}

Here, we compare the final converged energies obtained using ML-PQE and conventional PQE for a few molecular systems using dUCCSD ansatz (Figure. \ref{fig:accuracy}). In all of our calculations presented in this manuscript, only those double excitations are included in the ansatz, whose associated absolute amplitudes at the MP2 (M{\o}ller-Plesset perturbation theory to second order) level are greater than $10^{-05}$ . The parameters  ($\theta_{\mu}$), associated with operators ($\hat{\kappa}_{\mu}$) (see Eq. \eqref{6}), are initialized at their corresponding MP2 values for both conventional and ML aided methods. The convergence is marked by the \textit{2-norm} of the residue vector going below a threshold value of $10^{-5}$ for both methods. It is important to note that the residue vector for ML-PQE ansatz only includes residues corresponding to operators ($\hat{\kappa}_{\mu}$) that belong to \textit{PPS}. The value of the complexity parameter ($\alpha$) is kept at $10^{-10}$ and the LRNT is set to 0.007.

\begin{figure*}[!ht]
    \centering
\includegraphics[width=0.9\textwidth]{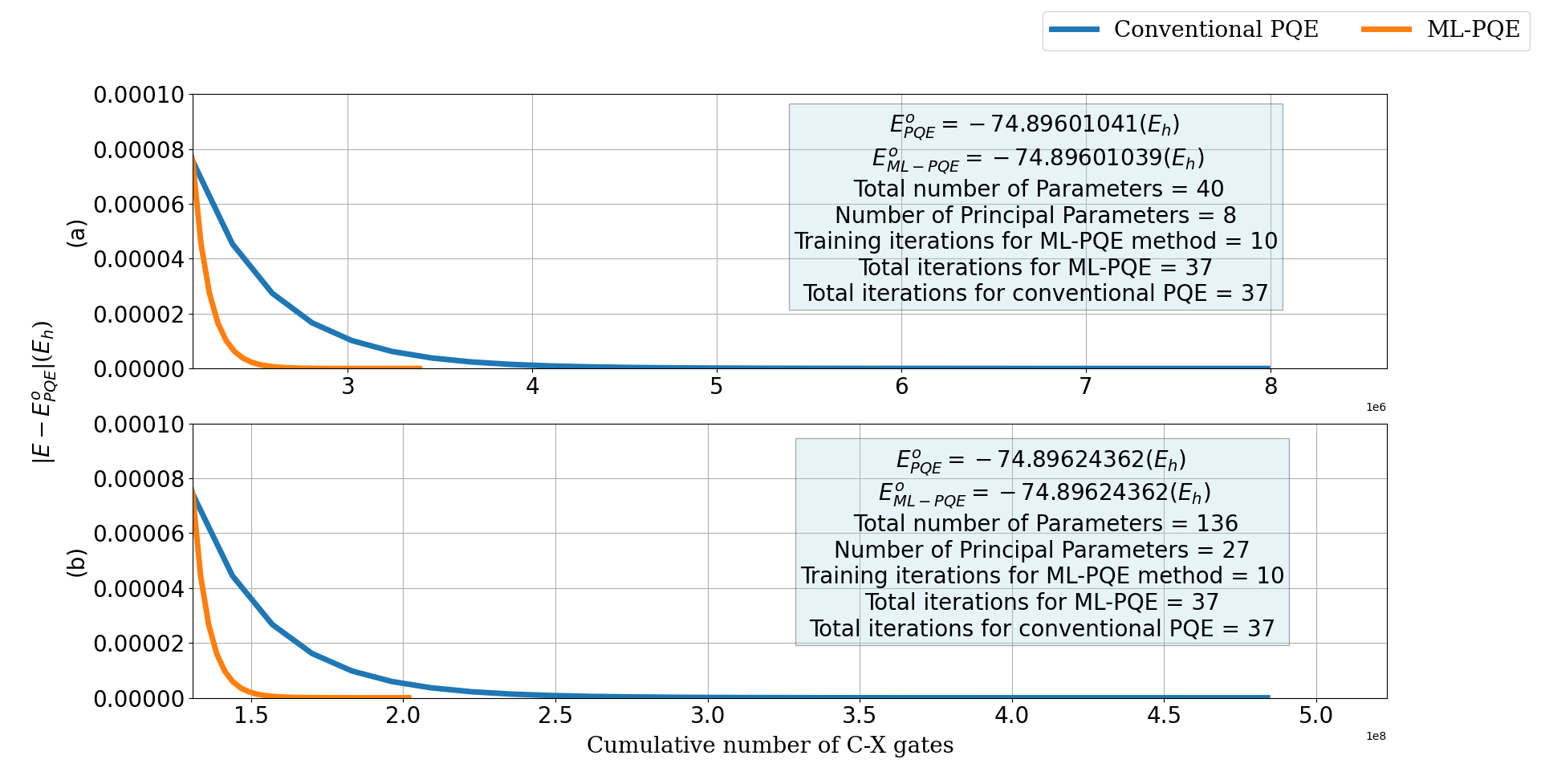}
\caption{Cumulative C-X gate utilization for convergence (in energy) trajectory for $H_2O$ at $1.5 \times r_{O-H}$ $(r_{O-H} = 0.958 \mbox{\normalfont\AA})$ using (a) dUCCSD and (b) dUCCSDT ansatz for ML-PQE and Conventional PQE algorithm. $E_{PQE}^o$ is the converged conventional PQE energy. E is the energy at each iteration for the respective methods.}
    \label{fig:cost}
\end{figure*}

As discernible from the graph presented in Figure \ref{fig:accuracy}, we obtain an energy difference of the order of $10^{-06}$ or lower
for nearly all of the scrutinized systems. 
While it is unreasonable to anticipate that ML-PQE will be exactly equivalent to the conventional method, the differences in the converged energies are negligible, and it is unlikely to produce any practical ramifications. 
Furthermore, it highlights the ability of the chosen ML model to recognize the relevant patterns in the training data and make meaningful predictions. The exemplary performance also confirms the validity of Eq. \eqref{functional generic rel.} and those leading to it in the context of PQE.
The overall number of iterations for ML-PQE, as indicated in the plot, encompasses the training iterations as well. 
Upon completion of training, the number of measurements performed per residue vector drops tremendously, as expressed by Eq. \ref{reduced_measurement}. Since this equation merely provides an upper limit, to demonstrate this reduction numerically, we exhibit an energy convergence plot against the cumulative controlled-X (C-X) gates (Fig. \ref{fig:cost}) utilized during the construction of the residue vectors post-training.
During the training phase, each ML-PQE iteration uses the same number of C-X gates as the conventional method.
However, when the ML mechanism is activated, the employment of C-X gates per iteration markedly declines, signifying the calculation of a reduced residue vector.

\begin{figure*}[!ht]
    \centering
\includegraphics[width=\textwidth]{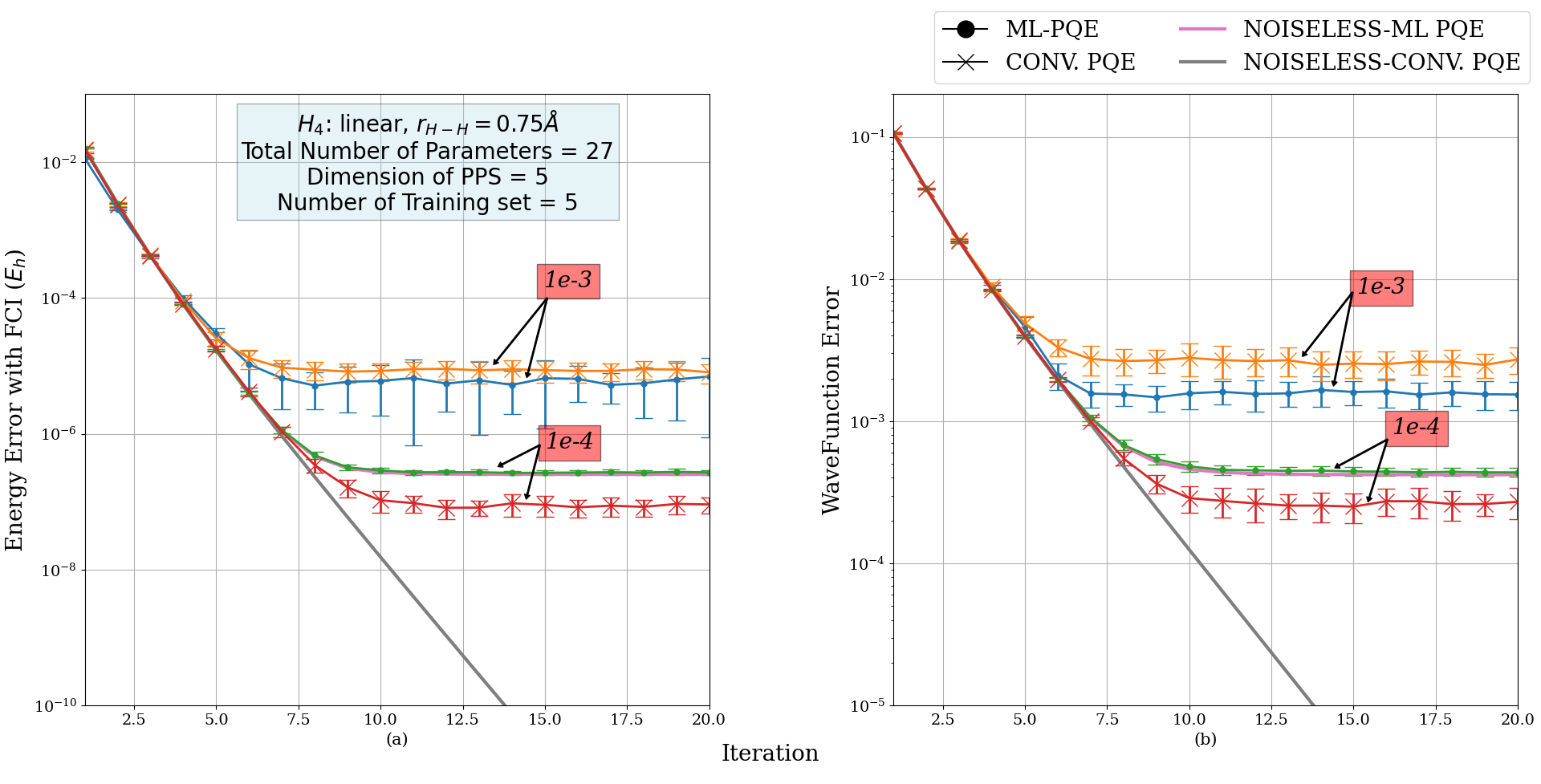}
    \label{fig:0.75}
\end{figure*}

\begin{figure*}[!ht]
    \centering
    \captionsetup{justification=centering}
\includegraphics[width=\textwidth]{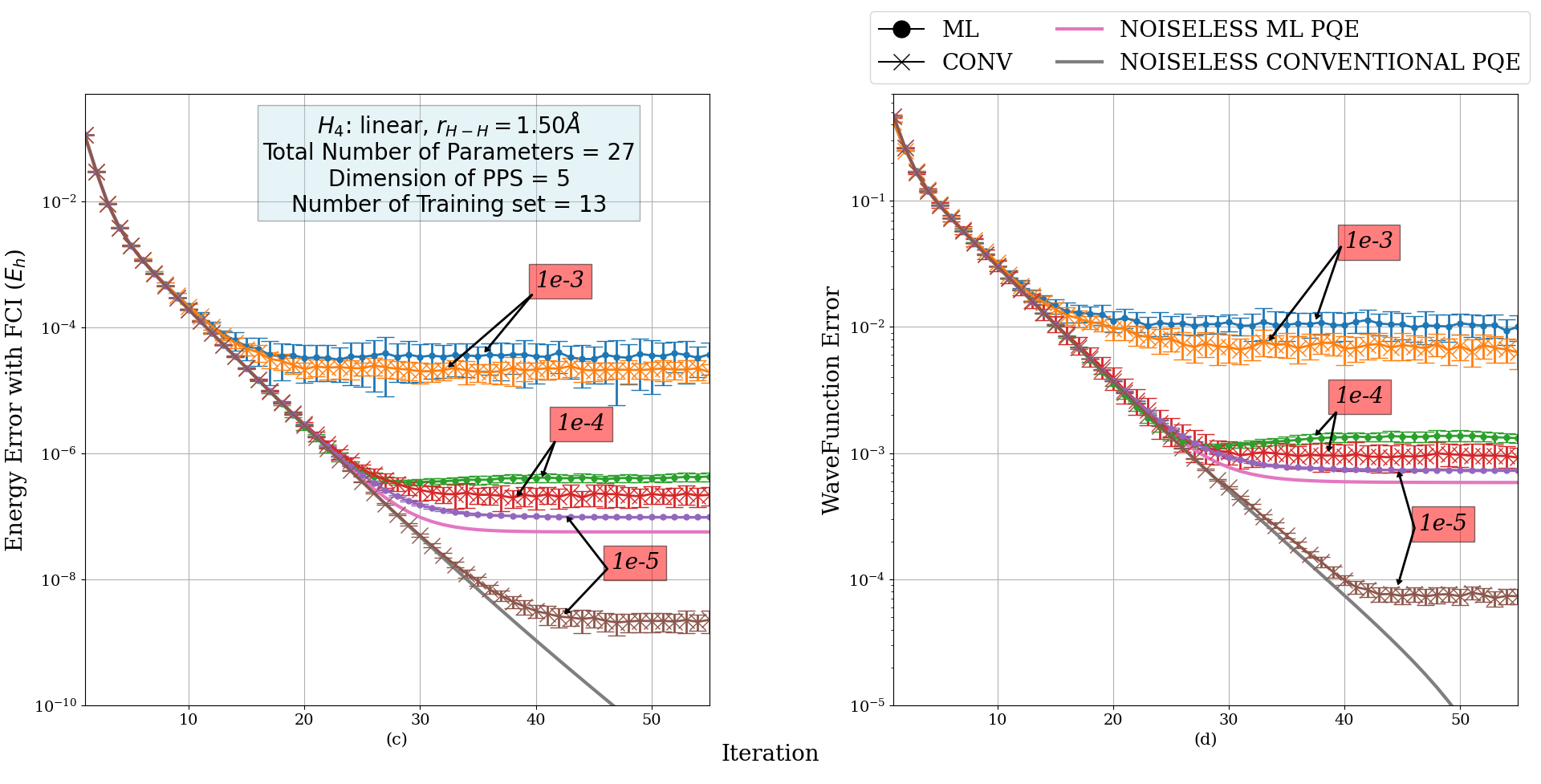}
\caption{Energy Error and Wave-Function Error (defined as $\|\theta_n-\theta_{NSL}^o\|$, where $\theta_n$ is the parameter vector at $n^{th}$ iteration and $\theta_{NSL}^o$ is the parameter vector obtained after the last iteration for the noiseless conventional PQE) for $H_{4}$ Linear chain: $r_{H-H}=0.75\mbox{\normalfont\AA}$(a,b) and $r_{H-H}=1.5\mbox{\normalfont\AA}$(c,d) at different noise level($\sigma$ denoted inside red boxes) at different iteration number. Each curve is obtained by averaging data of 50 calculations with error bars denoting the standard deviation. The training is done over averaged parameters. Number of training set = 5 for (a) and (b)(obtained by setting LRNT=0.02 at noiseless ML-PQE) and 13 for (c) and (d)(obtained by setting LRNT=0.005 at noiseless ML-PQE).$\alpha$ is taken as $10^{-10}$ at all noise levels in Fig (a) and (b) whereas for Figures (c) and (d), $\alpha$ is taken as $10^{-09}$ at noise level $\sigma=10^{-05}$ and increased to $10^{-06}$ to handle noise of $\sigma=10^{-04}$ and $\sigma=10^{-03}$  for ML-PQE calculations}
    \label{fig:1.5}
\end{figure*}

It could be contended that the usage of accelerating methodologies such as the DIIS within
conventional PQE could also result in a reduction in the cumulative measurements. However, it is vital to take into account that the stability of
DIIS is not always guaranteed, in contrast to the ML based approach, which operates seamlessly for all of the tested systems.
Instead of treating the two techniques as adversaries, one could potentially amalgamate DIIS with ML-based PQE to further enhance its cost
effectiveness. In our investigation, we have abstained from utilizing any such accelerators in order to conduct an impartial comparison and
accurately evaluate the performance of the ML-PQE.

\subsection{Effect of Stochastic Noise on ML-PQE}
The results presented in the preceding subsection were all obtained using the \textit{statevector} simulator, which represents a fault tolerant
quantum device with precise measurements. However, in the current state of quantum hardware, a multitude of noise-generating components can hamper
the accuracy of computations. Some of the major contributors to noise include unwanted interactions with the environment, errors in gate
realization, decoherence, and readout errors. It leads to inaccurate calculation of the residue elements which are used during the parameter update
equation. In this section, we study the manifestation of noise in the ML-PQE scheme and compare it with the conventional one. To imitate the effect
of noise, we add a random error ($\eta(0, \sigma)$), sampled from a Gaussian distribution centred at 0 and having a standard deviation of $\sigma$,
to all the residues calculated in a noiseless simulation ($r_{\mu}^{NSL}$). It is to be noted that this particular treatment of noise does not
capture more delicate and device dependent noise\cite{PRXQuantum.2.030301}. One can mathematically represent this as:
\begin{equation} \label{noise_residue}
    r_{\mu}^{M} = r_{\mu}^{NSL} + \eta(0, \sigma)
\end{equation}
 the sampling being done from a Gaussian distribution with the probability density given as:

\begin{equation}
    p(x) = \frac{1}{\sqrt{2\pi\sigma^2}} e^{\frac{-x^2}{2\sigma^2}}
\end{equation}

We perform ML-PQE and conventional PQE calculations with the residues affected by the noise (as mentioned in Eq. \eqref{noise_residue}) at different values of the standard deviation ($\sigma$) and compare them by plotting the energy error (with FCI) against iteration number for a linear chain $H_4$ molecule using a dUCCSDTQ ansatz at two geometries - $r_{H-H}=0.75 \mbox{\normalfont\AA}$ and $r_{H-H}=1.50 \mbox{\normalfont\AA}$ (Fig. \ref{fig:1.5}).

Upon inspection of Figure \ref{fig:1.5}, it is evident that, for a given level of noise strength characterized by the value of $\sigma$, the energy error for the ML-PQE is of a similar order as that of the conventional PQE. 
In general, noise is expected to impair the outcomes obtained using the ML-PQE scheme since it complicates the task of the ML model to recognize accurate patterns within the data.
It manifests in the form of overfitting, wherein the model tries to learn each data point diligently rather than finding a general trend. 
Different models come equipped with different inherent mechanisms to deal with this.
KRR uses regularization to combat overfitting, which can be accessed through the complexity parameter.
From Fig .\ref{fig:1.5}, one can see that the value of the complexity parameter had to be increased along with taking averaged data for training to keep the error in the same order (either above or below) as that of conventional PQE. As one decreases the strength of the noise (lowers $\sigma$), the trajectories corresponding to the ML-PQE and the conventional one approaches their noiseless marks where the former reaches its accuracy limit that is dictated by the performance of the ML model. 
Apart from the energy error, we have also plotted the \textit{wave function error} to give more insight into the activity of the parameters during the convergence. It follows the same trend as the energy error plots.

These plots showcase the ability of ML-PQE to gain noise resilience similar to the conventional one through regularization and establish the suitability of  ML-PQE for NISQ devices.

\section{Conclusions and Future Outlook}
Within the present endeavor, we have delved into the inherent characteristics that emerge from the non-linear iterative convergence protocol utilized in PQE, integrating this method with the slaving principle. Through this innovative approach, we have successfully established a functional relationship between the \textit{principal} and \textit{auxiliary} parameters, thereby enabling us to depict each iterative point in the phase space trajectory with a mere subset of parameters - the so-called \textit{PPS}. This substantial reduction in the degrees of freedom culminates in considerable savings in the residue computations conducted on a quantum device.

To capture the intricate inter-dependencies that exist between the members of the \textit{PPS} and \textit{APS}, we have resorted to the utilization of supervised ML techniques, specifically the Kernel Ridge Regression (KRR) model. In addition, we have also demonstrated the efficacy of this ML assisted PQE - the ML-PQE - under noisy simulation, where we have factored in a stochastic error in the residue. Moreover, the ML-PQE method is characterized by its independence from the order and size of the ansatz operator pool, and its applicability extends to any adaptively built ansatz.

As a final outcome of our efforts, we have succeeded in accelerating the iterative convergence protocol that is intrinsic to the solution of the non-linearly coupled PQE equations via the employment of ML. A rigorous examination of various ML models, coupled with hyper-parameter tuning for enhanced predictions and robust noise handling capabilities, would lay the groundwork for future research. Furthermore, it is worth noting that the analytical determination of the non-linear dependence of \textit{auxiliary} parameters on the \textit{principal} ones poses an extremely daunting and intricate challenge that must be met head-on by future researchers in this field.


\section{Acknowledgments}
The authors thank Mr. Samrendra Roy, IIT 
Bombay, for many stimulating discussions about the
details pertaining to ML techniques. RM acknowledges the 
financial support from Industrial Research and
Consultancy Centre, IIT Bombay, and 
Science and Engineering Research Board, Government
of India. SH thanks CSIR (Council of Scientific and Industrial Research), CP acknowledges UGC (University Grants Commission) and DM thanks PMRF (Prime Minister's Research Fellowship) for their respective research fellowships.

\section*{Data Availability}
The data is available upon reasonable request to the corresponding author.


\bibliography{./literature}

%

    
\end{document}